\documentclass[twocolumn,showkeys,aps,prb,showpacs]{revtex4-1}
\usepackage{graphicx}
\usepackage[CJKbookmarks,dvipdfm,colorlinks,linkcolor=blue,citecolor=blue]{hyperref}

\begin{document}

\title{Piezoelectric quantum spin Hall insulator with Rashba spin splitting  in Janus monolayer $\mathrm{SrAlGaSe_4}$}

\author{San-Dong Guo$^{1}$ Yu-Tong Zhu$^{1}$, Wen-Qi Mu$^{1}$ and  Xing-Qiu Chen$^{2,3}$}
\affiliation{$^1$School of Electronic Engineering, Xi'an University of Posts and Telecommunications, Xi'an 710121, China}

\affiliation{$^2$Shenyang National Laboratory for Materials Science, Institute of Metal Research,
Chinese Academy of Science, 110016 Shenyang, Liaoning, P. R. China}
\affiliation{$^3$School of Materials Science and Engineering, University of Science and Technology of China,
Shenyang 110016, P. R. China}
\begin{abstract}
The realization of multifunctional two-dimensional (2D) materials is
fundamentally intriguing, such as combination of piezoelectricity with   topological
insulating phase or  ferromagnetism.
In this work, a Janus monolayer $\mathrm{SrAlGaSe_4}$ is built from 2D $\mathrm{MA_2Z_4}$ family  with  dynamic, mechanical and thermal  stabilities, which is piezoelectric due to lacking inversion symmetry.
The unstrained $\mathrm{SrAlGaSe_4}$ monolayer is a narrow gap normal insulator (NI) with spin orbital coupling (SOC).
However, the NI to topological insulator (TI) phase transition can be induced by the biaxial strain, and a piezoelectric quantum spin Hall insulator (PQSHI) can be achieved. More excitingly, the phase transformation point is only about 1.01 tensile strain, and nontrivial band topology can hold until considered 1.16 tensile strain. Moreover, a Rashba spin splitting in the conduction bands can exit in PQSHI due to  the absence of a
horizontal mirror symmetry and the presence of SOC. For monolayer $\mathrm{SrAlGaSe_4}$, both  in-plane and much weak out-of-plane piezoelectric polarizations can be induced with a uniaxial strain  applied.
The calculated piezoelectric strain  coefficients $d_{11}$ and $d_{31}$ of  monolayer $\mathrm{SrAlGaSe_4}$ are  -1.865 pm/V and -0.068 pm/V at 1.06 tensile strain as a representative TI. In fact, many  PQSHIs can be realized from 2D $\mathrm{MA_2Z_4}$ family. To confirm that,  similar to $\mathrm{SrAlGaSe_4}$, the coexistence of  piezoelectricity and  topological orders can be realized by strain (about 1.04 tensile strain) in the $\mathrm{CaAlGaSe_4}$ monolayer.
Our works  suggest that  Janus monolayer $\mathrm{SrAlGaSe_4}$ is a pure 2D system for PQSHI, enabling future studies exploring the interplay between piezoelectricity and topological orders,  which can lead  to novel applications in electronics and spintronics.

\end{abstract}

\pacs{71.20.-b, 77.65.-j, 72.15.Jf, 78.67.-n ~~~~~~~~~~~~~~~~~~~~~~~~~~~~~~~~~~~Email:sandongyuwang@163.com}

\maketitle

\section{Introduction}
The multifunctional 2D materials can provide a unique opportunity
for intriguing physics and practical device applications.
 The combination of piezoelectricity and ferromagnetism   is one
such example, which has been predicted in 2D vanadium dichalcogenides and septuple-atomic-layer  $\mathrm{VSi_2P_4}$\cite{qt1,q15}.
Other example is  ferroelastic TI or  ferroelectric TI, which  simultaneously possesses ferroelastic or ferroelectric and  quantum spin Hall (QSH) characteristics.  The  2D Janus TMD MSSe (M = Mo and W) monolayers have been predicted as ferroelastic TI\cite{qt2}, and the  $\mathrm{In_2Te_3}$/$\mathrm{In_2Se_3}$ bilayer heterostructure as ferroelectric TI\cite{qt2-1}.
Antiferromagnetic TI with coexistence of intrinsic antiferromagnetism and QSH states also has been reported in XMnY (X=Sr and Ba; Y=Sn and Pb) quintuple layers\cite{qt2-2}, as promising candidates for innovative spintronics applications. Piezoelectricity and band topology are two extensively studied
distinct properties of insulators, and  their coexistence  in a single 2D material may lead to novel physical phenomenon and device applications.

 The piezoelectric 2D semiconducting material, allowing for energy conversion
between electrical and mechanical energy, has been a research
focus in the ever-increasing energy conversion area\cite{q4,q4-1}.
A 2D semiconducting material can
 show piezoelectric properties,  when inversion symmetry is broken.
A typical 2D piezoelectric material is  monolayer $\mathrm{MoS_2}$ with 2H phase, which  is firstly   predicted by the first principles calculations \cite{q11}, and then is confirmed in experiment\cite{q5,q6}.
In theory,  many 2D monolayers  are predicted as  potential  2D piezoelectric materials by density functional theory (DFT) calculations\cite{q11,q7,q9,q10,q12,q13,q14,q15,q16,q17}. Piezoelectricity can be related with valley chern number in inhomogeneous
hexagonal 2D crystals\cite{q17-1}, and nonlinear exciton drift in piezoelectric 2D materials has also been reported\cite{q17-2}.

For TI with spin-momentum-locked conducting
edge states and insulating properties in the bulk, the charge and spin transport in the
edge states are quantized dissipationless, endowing them
with  rich physics and
promising applications in spintronics and quantum computations\cite{t1,t2}.
A 2D TI is also called as QSH insulator (QSHI) for its quantized edge conductance, and graphene is  the first predicted 2D TI
characterized by counter-propagating edge currents with
opposite spin polarization\cite{t3}.
Experimentally, the  HgTe/CdTe and InAs/GaSb quantum wells  have been confirmed as QSHIs \cite{t4,t5}, and  lots of  QSHIs
 have also  been theoretically proposed by DFT calculations\cite{t6,t7,qt3,t8,t9,t10}.

\begin{figure*}
  \includegraphics[width=16.0cm]{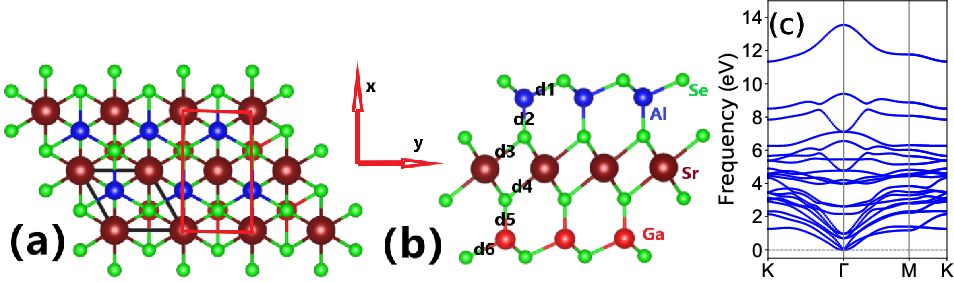}
  \caption{(Color online)The (a) top view and (b) side view of crystal structure of monolayer $\mathrm{SrAlGaSe_4}$, and   the  rhombus primitive cell  and   rectangle supercell are marked  by  black and red frames, along with bond lengths $d_i$. (c)The phonon band dispersions  of  $\mathrm{SrAlGaSe_4}$ monolayer by using GGA. }\label{t0}
\end{figure*}
\begin{figure}
  \includegraphics[width=8cm]{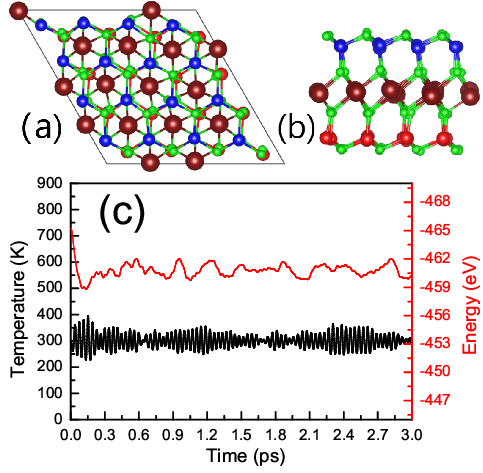}
\caption{(Color online) The (a) top view and (b) side view of  crystal structures  of $\mathrm{SrAlGaSe_4}$  monolayer after the simulation for 3 ps at 300 K.  (c):the  temperature and total energy fluctuations of $\mathrm{SrAlGaSe_4}$  monolayer at 300 K.}\label{md}
\end{figure}

Coexistence of intrinsic piezoelectricity and nontrivial band topology has been predicted  in monolayer  InXO (X=Se and Te) from our previous works\cite{gsd}.
However,  the   InXO (X=Se and Te) monolayers are built from monolayer InX (X=Se and Te)  by
oxygen functionalization with chemisorption of oxygen atoms on  both sides. It may be more practical to search for pure PQSHI, which can easily be confirmed in experiment. Recently, the 2D $\mathrm{MA_2Z_4}$ family has attracted a surge of interest\cite{msn,m20,m21,m22,m23,q13,q14,m26}, and the $\mathrm{MoSi_2N_4}$ and $\mathrm{WSi_2N_4}$ monolayers are successfully synthesized by chemical vapor deposition (CVD)\cite{msn}.
Intrinsic piezoelectricity  has been found in  $\mathrm{MA_2Z_4}$ family\cite{q15,q13,q14}, such as $\alpha_1$-$\mathrm{MoSi_2N_4}$ and $\mathrm{WSi_2N_4}$.
On the other hand, $\beta_2$-$\mathrm{SrGa_2Se_4}$ is predicted to be a 2D TI with generalized gradient approximation (GGA)\cite{m20}.
It's a natural idea to achieve PQSHI in the new septuple-atomic-layer 2D $\mathrm{MA_2Z_4}$ family.

In this work,  we propose a design principle for the realization
of PQSHI in 2D $\mathrm{MA_2Z_4}$ family. Firstly, we chose a 2D TI $\beta_2$-$\mathrm{SrGa_2Se_4}$ from  $\mathrm{MA_2Z_4}$ family, which is centrosymmetric, lacking piezoelectricity. Secondly, to realize piezoelectric response,  the inversion symmetry
is broken by constricting Janus $\mathrm{SrAlGaSe_4}$ monolayer, which can be attained by  replacing the Ga atoms of  top GaSe bilayer in  $\mathrm{SrGa_2Se_4}$  monolayer with Al atoms. Finally, the biaxial  strain is used to tune the topological properties of  $\mathrm{SrAlGaSe_4}$ monolayer. The $Z_2$ topological invariant is used to recognize the nontrivial topological state. Interestingly, by very small strain (about 1.01 tensile strain), the $\mathrm{SrAlGaSe_4}$ monolayer can become PQSHI with additional Rashba spin splitting. To further confirm our design principle,
 similar to $\mathrm{SrAlGaSe_4}$, the PQSHI can be realized  in the $\mathrm{CaAlGaSe_4}$ monolayer by about 1.04 tensile strain.
 Therefore, the  coupling between topological state and piezoelectricity is identified, offering a kind of new multifunctional 2D materials  for novel designs in spintronics or
optoelectronics.

The rest of the paper is organized as follows. In the next
section, we shall give our computational details and methods.
 In  the next few sections,  we shall present crystal structure and stability, electronic structures and  piezoelectric properties of Janus  monolayer  $\mathrm{SrAlGaSe_4}$.  Finally, we shall give our discussion and conclusions.

\begin{table}
\centering \caption{For monolayer $\mathrm{SrAlGaSe_4}$, the lattice constants $a_0$ ($\mathrm{{\AA}}$), the bond lengths $d_i$ ($\mathrm{{\AA}}$), the elastic constants $C_{ij}$ ($\mathrm{Nm^{-1}}$), shear modulus
$G_{2D}$ ($\mathrm{Nm^{-1}}$),  Young's modulus $C_{2D}$  ($\mathrm{Nm^{-1}}$),  Poisson's ratio $\nu$,  the GGA and GGA+SOC gaps  (meV). }\label{tab0}
  \begin{tabular*}{0.48\textwidth}{@{\extracolsep{\fill}}ccccccc}
  \hline\hline
$a_0$&$d_1$ &$d_2$&$d_3$&$d_4$&$d_5$&$d_6$\\\hline
4.07 &2.56&2.30&3.00&2.99&2.33&2.58\\\hline\hline
$C_{11}$ &  $C_{12}$& $G_{2D}$&$C_{2D}$& $\nu$& $Gap$& $Gap_{SOC}$\\\hline
82.67&30.14&26.27 & 71.68&0.37&89.8&12.9\\\hline\hline
\end{tabular*}
\end{table}

\section{Computational detail}
Within DFT\cite{1}, the first-principles calculations  are carried out using  the projected augmented wave
(PAW) method, as implemented in
the  VASP package\cite{pv1,pv2,pv3}.  The total energy  convergence criterion is set
to $10^{-8}$ eV with the cutoff energy
for plane-wave expansion being 500 eV. We use GGA of Perdew, Burke and  Ernzerhof  (GGA-PBE)\cite{pbe}  as the exchange-correlation potential, and the SOC is considered to investigate electronic structures and piezoelectric stress coefficients $e_{ij}$. A Monkhorst-Pack mesh of 16$\times$16$\times$1 is adopted for geometry optimization  with the residual force on each atom being less than 0.0001 $\mathrm{eV.{\AA}^{-1}}$.
The vacuum region along the z direction is set to more than 20 $\mathrm{{\AA}}$ in order to
decouple
the spurious interaction  between
the layers. The constant energy contour plots of the spin
texture are calculated by the PYPROCAR code\cite{py}.

The  Phonopy code\cite{pv5} is used  to calculate  phonon dispersion
spectrums of studied monolayers with a supercell
of 5$\times$5$\times$1 by finite displacement method. A 2$\times$2$\times$1 k-mesh is employed with kinetic energy cutoff of 500 eV to calculate the second order interatomic force constants (IFCs).
To obtain the piezoelectric strain coefficients $d_{ij}$, the elastic stiffness tensor $C_{ij}$ and  piezoelectric stress coefficients $e_{ij}$ are calculated by using  strain-stress relationship (SSR) and density functional perturbation theory (DFPT) method\cite{pv6}.
The 2D elastic coefficients $C^{2D}_{ij}$
 and   piezoelectric stress coefficients $e^{2D}_{ij}$
have been renormalized by the the length of unit cell along z direction ($Lz$):  $C^{2D}_{ij}$=$Lz$$C^{3D}_{ij}$ and $e^{2D}_{ij}$=$Lz$$e^{3D}_{ij}$.
A Monkhorst-Pack mesh of 16$\times$16$\times$1 is adopted to calculate  $C_{ij}$ by GGA, and  5$\times$10$\times$1 for $e_{ij}$ by GGA+SOC.
A tight-binding Hamiltonian with the
maximally localized Wannier functions is constructed to fit  band structures from the first-principles calculations, and then the $Z_2$ invariants are calculated, as implemented in the Wannier90 and WannierTools
codes\cite{w1,w2}.

\begin{figure}
  \includegraphics[width=8cm]{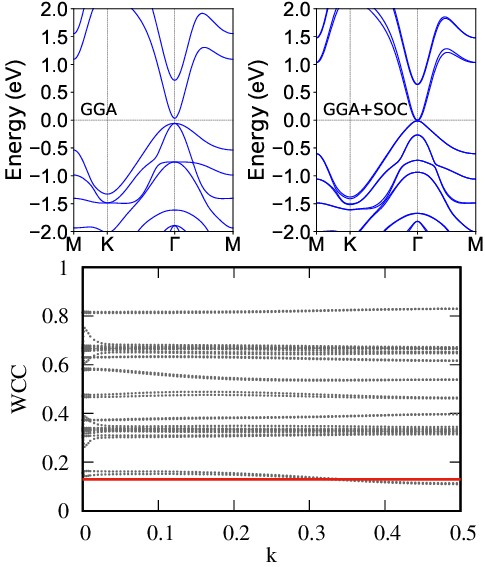}
\caption{(Color online)Top:The energy band structures  of  $\mathrm{SrAlGaSe_4}$ monolayer  using GGA  and GGA+SOC. Bottom:Evolution of WCC of $\mathrm{SrAlGaSe_4}$ monolayer at unstrained condition }\label{band}
\end{figure}

\begin{figure*}
   \includegraphics[width=16cm]{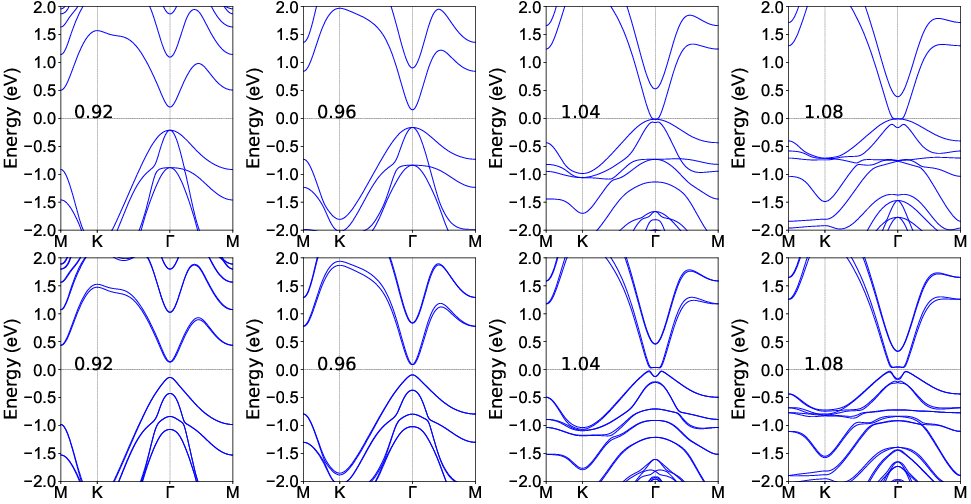}
\caption{(Color online) The energy band structures  of $\mathrm{SrAlGaSe_4}$ monolayer  using GGA (Top) and GGA+SOC (Bottom) at representative 0.92, 0.94, 1.04 and 1.08 strains.}\label{band-s}
\end{figure*}
\begin{figure}
   \includegraphics[width=8.0cm]{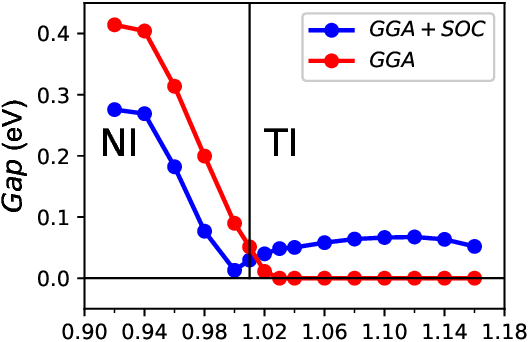}
  \caption{(Color online) The energy band gaps of $\mathrm{SrSiGeN_4}$ monolayer as a function of  $a/a_0$ by using both GGA and GGA+SOC.}\label{t3-1}
\end{figure}

\begin{figure*}
  \includegraphics[width=12cm]{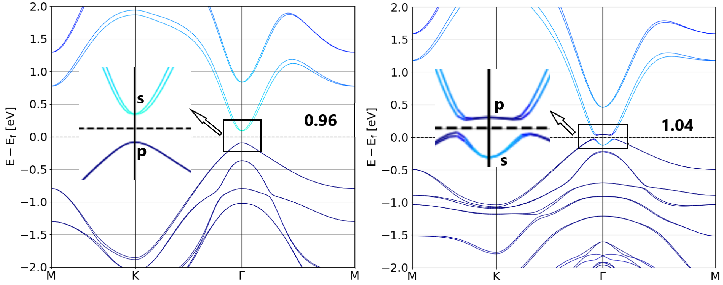}
\caption{(Color online)  The $s$ and $p$ orbital projected band structures of $\mathrm{SrAlGaSe_4}$ monolayer by using GGA+SOC at 0.96 and 1.04 strains. }\label{spp}
\end{figure*}
\begin{figure*}
  \includegraphics[width=12cm]{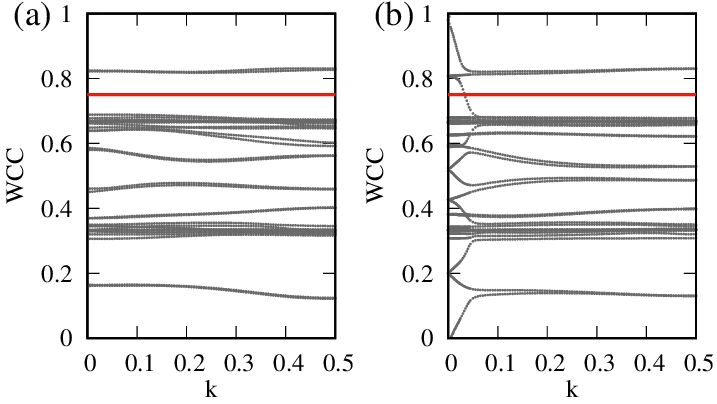}
\caption{(Color online)Evolution of WCC of $\mathrm{SrAlGaSe_4}$ monolayer at 0.96 (a) and 1.04 (b) strains. The $Z_2$=0 at 0.96 strain for topological trivial feature and $Z_2$=1 at 1.04 strain for topological non-trivial one.}\label{z2}
\end{figure*}

\section{Crystal structure and stability}
With $\mathrm{SrSe_2}$ triple layers sandwiched between the GaSe bilayers, the septuple-atomic-layer $\mathrm{SrGa_2Se_4}$ can be built with centrosymmetry and vertical reflection symmetry\cite{m20}, which has disappeared piezoelectricity. To stimulate  piezoelectricity, a natural way is to construct Janus structure, which can be achieved by  replacing the Ga atoms of  top GaSe bilayer in  $\mathrm{SrGa_2Se_4}$  monolayer with Al atoms, namely $\mathrm{SrAlGaSe_4}$ monolayer. The symmetry space group  of Janus monolayer $\mathrm{SrAlGaSe_4}$ is reduced to No.156 from No.164 of $\mathrm{SrGa_2Se_4}$ monolayer, which can induce  both in-plane and out-of-plane piezoelectricity.
The geometric structure of $\mathrm{SrAlGaSe_4}$ monolayer is shown in \autoref{t0}, along with  both rhombus primitive cell  and rectangle supercell.
With armchair and zigzag directions   defined as x and y directions, the rectangle supercell can be used to calculate piezoelectric coefficients $e_{ij}$. The optimized lattice
constant of the Janus $\mathrm{SrAlGaSe_4}$ monolayer is 4.07 $\mathrm{{\AA}}$, and the
interval distance between the upper Se layer and the lower Se layer is 10.38 $\mathrm{{\AA}}$. In $\mathrm{SrGa_2Se_4}$ monolayer, the equivalent  bond lengths between $d_1$ and $d_6$, or $d_2$ and $d_5$, or $d_3$ and $d_4$ can be observed. For monolayer $\mathrm{SrAlGaSe_4}$,
the difference in atomic sizes and electronegativities of Al and Ga atoms leads to inequivalent  bond lengths and
bond characteristics, which can be observed from \autoref{tab0}. The inequivalent  bond lengths and
bond characteristics can induce an electrostatic potential gradient, and then built-in electric field can be attained, which can give rise to Rashba spin splitting.

To study the stability of  Janus $\mathrm{SrAlGaSe_4}$ monolayer, we firstly
calculate the phonon dispersion  to validate its dynamic stability. As plotted in
\autoref{t0}, there is no imaginary vibrational frequency with three acoustic
and eighteen optical phonon branches, which clearly suggests that $\mathrm{SrAlGaSe_4}$ monolayer is
dynamically stable. Moreover,  both
linear and flexural modes occur around the $\Gamma$ point, which can be observed in most 2D materials\cite{m20,gsd1,gsd2}.
In addition,  by performing ab-initio molecular dynamics (AIMD) simulations, we examine the thermal stability  of $\mathrm{SrAlGaSe_4}$ monolayer with a supercell of  4$\times$4$\times$1 for more than 3000 fs  at 300 K. The temperature and total energy fluctuations of $\mathrm{SrAlGaSe_4}$ monolayer as a function of simulation time are plotted in \autoref{md}, along with the  crystal structures of $\mathrm{SrAlGaSe_4}$ at  300 K after the simulation for 3 ps. It is found that monolayer $\mathrm{SrAlGaSe_4}$ undergoes  no structural
reconstruction with  small temperature and total energy
 fluctuates around 300 K, which  indicates  the thermal stability of $\mathrm{SrAlGaSe_4}$ monolayer.

 To further check
the mechanical stability of $\mathrm{SrAlGaSe_4}$ monolayer,
the elastic constants $C_{ij}$ are calculated. Using Voigt notation, the elastic tensor with hexagonal symmetry can be given:
\begin{equation}\label{pe1-4}
   C=\left(
    \begin{array}{ccc}
      C_{11} & C_{12} & 0 \\
     C_{12} & C_{11} &0 \\
      0 & 0 & (C_{11}-C_{12})/2 \\
    \end{array}
  \right)
\end{equation}
The  two  independent elastic
constants of monolayer $\mathrm{SrAlGaSe_4}$  are $C_{11}$=82.67 $\mathrm{Nm^{-1}}$ and $C_{12}$=30.14 $\mathrm{Nm^{-1}}$.
The shear modulus is $G^{2D}$=26.27 $\mathrm{Nm^{-1}}$, which can be attained by ($C_{11}$-$C_{12}$)/2, namely $C_{66}$. The  calculated $C_{11}$ and  $C_{66}$  satisfy  the  Born  criteria of mechanical stability of a material with hexagonal symmetry \cite{ela}: $C_{11}$$>$0 and  $C_{66}$$>$0, which confirms  the mechanical stability of monolayer  $\mathrm{SrAlGaSe_4}$.
The Young's modulus $C_{2D}$ are given\cite{ela1}:
\begin{equation}\label{e1}
C_{2D}=\frac{C_{11}^2-C_{12}^2}{C_{11}}
\end{equation}
The calculated $C_{2D}$ is 71.68 $\mathrm{Nm^{-1}}$, which are very smaller  than ones of  monolayer  $\mathrm{MSi_2N_4}$ (M=Ti, Zr, Hf, Cr, Mo and W) and Janus  $\mathrm{MSiGeN_4}$ (M=Mo and W)\cite{q13,gsd1}, indicating that  monolayer $\mathrm{SrAlGaSe_4}$ is not rigid.
The Poisson's ratio $\nu$ is also calculated by $C_{12}$/$C_{11}$, and it is 0.37.
The phonon calculations, AIMD and
elastic constants show dynamical, thermal  and mechanical  stability of the monolayer $\mathrm{SrAlGaSe_4}$, suggesting its  possible synthesis.

\begin{figure*}
  \includegraphics[width=12cm]{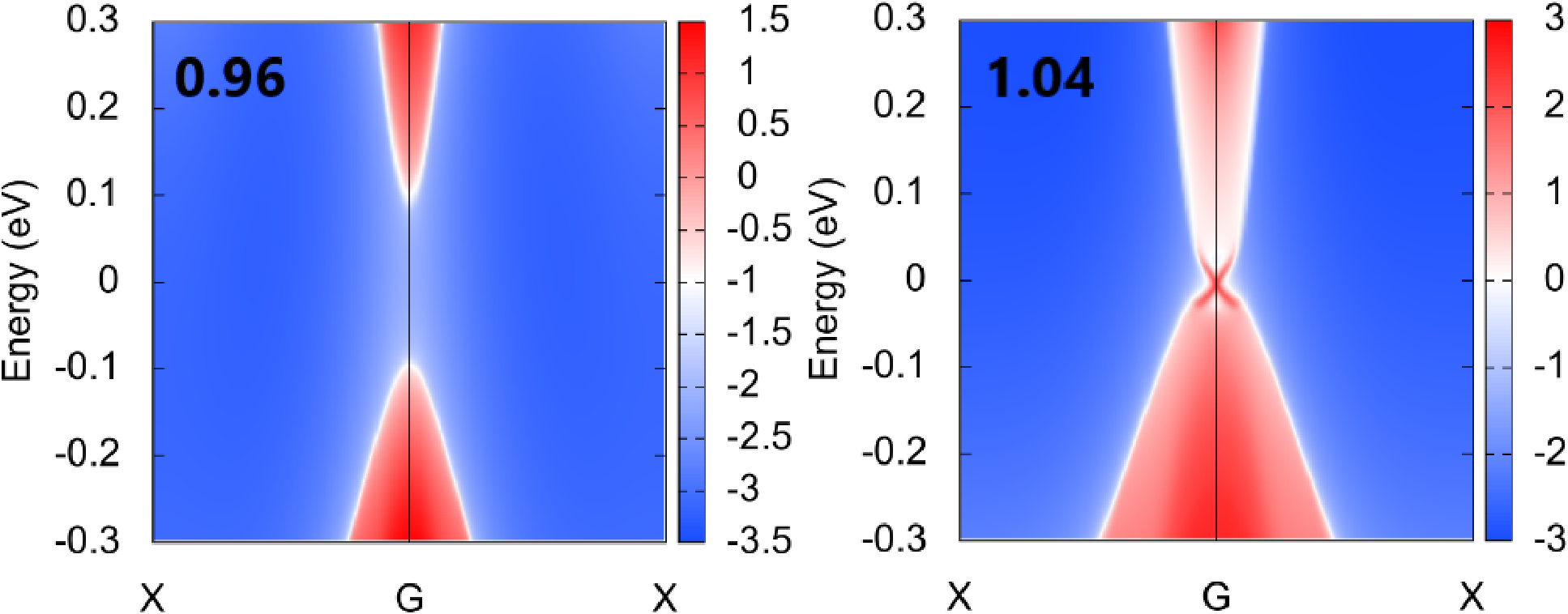}
\caption{(Color online) The edge states of semi-infinite $\mathrm{SrAlGaSe_4}$ under 0.96 and 1.04 strains. }\label{ss}
\end{figure*}
\begin{figure*}
  \includegraphics[width=15cm]{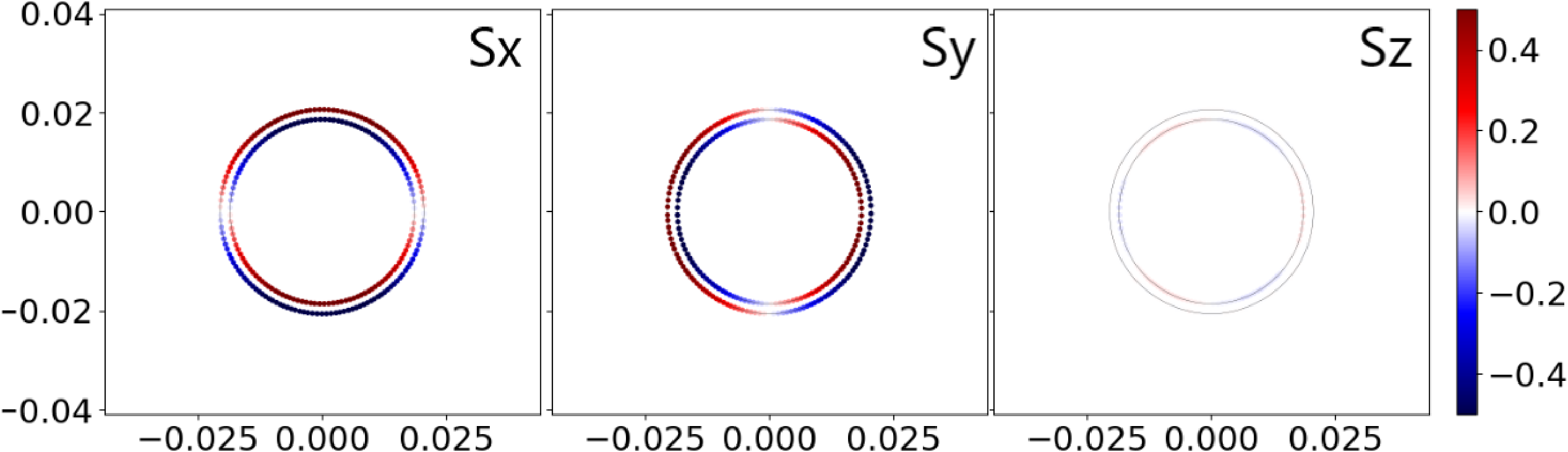}
\caption{(Color online)  Spin texture calculated in  $k_z=0$ plane centered at the $\Gamma$ point at the iso-energy surface of 0.2
eV above the Fermi level for  $\mathrm{SrAlGaSe_4}$ monolayer. The red and blue colours show spin-up and spin-down states, respectively.}\label{rs}
\end{figure*}

\section{Electronic structures}
 With GGA and GGA+SOC, the energy bands of monolayer $\mathrm{SrAlGaSe_4}$  are shown in \autoref{band}.
The GGA results show that the  monolayer $\mathrm{SrAlGaSe_4}$ is direct gap semiconductor (89.8 meV) with the valence band maximum (VBM) and conduction band minimum (CBM) being at the $\Gamma$ point. When including SOC, a indirect gap of 12.9 meV is observed, and the CBM still locates at $\Gamma$ point, but the VBM deviates slightly from $\Gamma$ point. The other  SOC effects are that the Rashba spin splitting near the CBM  and the
spin-orbit splitting of 209 meV  at VBM  are observed. It is found that the conduction and valence bands near the Fermi level are mainly $s$ and $p$
dominated ones, respectively. It has been proved that monolayer $\mathrm{SrGa_2Se_4}$ is a 2D TI by using GGA\cite{m20}.
In order to ascertain the topological properties
in the monolayer  $\mathrm{SrAlGaSe_4}$, we calculate the $Z_2$
topological invariants, which can be confirmed via calculations of the Wannier charge center (WCC).   If   $Z_2$ equals 1, a material is  a topologically nontrivial state, and  $Z_2$ = 0 means a trivial state. According to WCC in \autoref{band}, it is clearly seen that the number of crossings between the WCC and the reference horizontal line is even, which means that monolayer $\mathrm{SrAlGaSe_4}$  is NI.

To achieve PQSHI, a strategy should be used to realize topological state in  monolayer $\mathrm{SrAlGaSe_4}$.
Although monolayer $\mathrm{SrGa_2Se_4}$ and $\mathrm{SrAl_2Se_4}$ have the same configuration of outer shell electrons, the $\mathrm{SrGa_2Se_4}$ is TI, while $\mathrm{SrAl_2Se_4}$ is NI\cite{m20}. The possible reason is that $\mathrm{SrGa_2Se_4}$ has larger lattice constants than $\mathrm{SrAl_2Se_4}$. The lattice constants of  $\mathrm{SrAlGaSe_4}$  is between ones of monolayer $\mathrm{SrGa_2Se_4}$ and $\mathrm{SrAl_2Se_4}$. So, it is possible to achieve topologically nontrivial state in  monolayer $\mathrm{SrAlGaSe_4}$ by strain engineering. Similar idea has been used in monolayer BiSb and SbAs, and NI to TI transition can be induced by  biaxial
tensile strain\cite{zsl,zsl1}. Here, we use $a/a_0$ to simulate compressive/tensile strain  with $a$ and $a_0$ being the strained and  unstrained lattice constants, respectively. The  $a/a_0$$<$1/$a/a_0$$>$1 means compressive/tensile strain. The strain range from 0.90 to 1.16 is considered to calculate the electronic structures of  $\mathrm{SrAlGaSe_4}$ monolayer. The energy bands at representative strain points are plotted in \autoref{band-s} with both GGA and GGA+SOC, and the energy band gaps of both GGA and GGA+SOC as $a$/$a_0$ function are shown in \autoref{t3-1}.
From 0.90 to 1.03 strain, the GGA gap decreases, and then the GGA gap is always zero from 1.03 to 1.16 strain.
 When  including SOC, the SOC effect  opens  gap from 1.03 to 1.16 strain, which suggests that  monolayer $\mathrm{SrAlGaSe_4}$ may become
 potential 2D TI. However, we calculate $Z_2$ at all strain points to confirm critical point of NI to TI, and it is about at 1.01, which means that
 very small tensile strain can induce NI to TI transition.

To further understand the NI and TI, the $s$ and $p$ orbital projected band structures of $\mathrm{SrAlGaSe_4}$ monolayer by using GGA+SOC at 0.96 and 1.04 strains are plotted in \autoref{spp}.
At 0.96 strain, the  CBM of $\mathrm{SrAlGaSe_4}$ at $\Gamma$ point comprises with  $s$-dominated orbits,
and its VBM consists of $p$-dominated orbits. In contrast, at 1.04 strain,  an opposite situation is observed,
and  the conduction bands near the Fermi level around $\Gamma$ point have  the $p$-dominated orbits, whereas the valence bands  now become
 $s$-dominated orbits, which means the occurrence of the electronic band inversion, implying the possible topological non-trivial feature.
 \autoref{z2} shows the WCCs of 0.96 and 1.04 strained $\mathrm{SrAlGaSe_4}$ monolayers. For an arbitrary horizontal reference line
(e.g. WCC=0.75), it crosses the evolution of WCC even number at 0.96 strain  and odd number at 1.04 strain, respectively.
In other words, the  $\mathrm{SrAlGaSe_4}$ monolayer at 0.96 strain is a trivial NI with $Z_2$=0, while under
1.04 tensile strain it transforms to a TI  with $Z_2$=1.  Furthermore, a TI has to exhibit non-trivial topological
edge states. The Green's-function method is used  to calculate the surface states on (100) surface based on the tight-binding
Hamiltonian, which are plotted in \autoref{ss} at 0.96 and 1.04 strains.
For the trivial NI under 0.96 compressive strain, no edge states  are observed. In contrast, for the TI under 1.04 tensile strain, topological helical edge states with the appearance of the Dirac cone is observed, which connect the conduction and valence bands.

Coexistence of intrinsic piezoelectricity and nontrivial band topology (namely PQSHI) in monolayer $\mathrm{SrAlGaSe_4}$ has been achieved by strain. Moreover,
the Rashba spin splitting can exist in PQSHT due to the breaking of vertical reflection symmetry. To examine the
Rashba effect,  the in-plane spin-texture of monolayer $\mathrm{SrAlGaSe_4}$ is calculated.
The spin projected constant energy contour plots (0.2
eV above the Fermi level) of the spin textures calculated in $k_x$-$k_y$
plane centered at the $\Gamma$ point are shown in \autoref{rs}.
For both $S_x$ and $S_y$ spin components, the pair of spin-splitting bands  have opposite spin orientation.
 The concentric spin-texture circles mean the purely 2D Rashba spin splitting at the conduction bands near the Fermi level. According to the projection
of different spin components, the  only in-plane $S_x$ and $S_y$ spin components are
present in the Rashba spin split bands, while  out-of-plane $S_z$ component disappears.
The in-plane spin moments  at the two rings have opposite
chirality with clockwise for the large ring and anticlockwise for the small ring, respectively.

\section{Piezoelectric properties}
 A noncentrosymmetric material with  applied strain or stress will induce  electric
dipole moments, which can produce an electrical voltage. The pristine  $\mathrm{SrGa_2Se_4}$ monolayer is non-piezoelectric due to
having centrosymmetry. However, monolayer $\mathrm{SrAlGaSe_4}$ with particular Janus structure will possess piezoelectric effect.
The piezoelectric response of a material can be described by  third-rank piezoelectric stress tensor  $e_{ijk}$ and strain tensor $d_{ijk}$, which  from the sum of ionic and electronic contributions are defined as:
 \begin{equation}\label{pe0}
      e_{ijk}=\frac{\partial P_i}{\partial \varepsilon_{jk}}=e_{ijk}^{elc}+e_{ijk}^{ion}
 \end{equation}
and
 \begin{equation}\label{pe0-1}
   d_{ijk}=\frac{\partial P_i}{\partial \sigma_{jk}}=d_{ijk}^{elc}+d_{ijk}^{ion}
 \end{equation}
In which $P_i$, $\varepsilon_{jk}$ and $\sigma_{jk}$ are polarization vector, strain and stress, respectively.
The superscripts $elc$ and $ion$ mean electronic and ionic contributions with $e_{ijk}^{elc}$/$d_{ijk}^{elc}$  ($e_{ijk}$/$d_{ijk}$) being clamped-ion (relax-ion) piezoelectric coefficients.
The $e_{ijk}$ and $d_{ijk}$ can be related by elastic tensor $C_{mnjk}$:
 \begin{equation}\label{pe0-1-1}
    e_{ijk}=\frac{\partial P_i}{\partial \varepsilon_{jk}}=\frac{\partial P_i}{\partial \sigma_{mn}}.\frac{\partial \sigma_{mn}}{\partial\varepsilon_{jk}}=d_{imn}C_{mnjk}
 \end{equation}

For 2D materials, only the in-plane strain and stress are taken into account (namely $\varepsilon_{jk}$=$\sigma_{ij}$=0 for i=3 or j=3)\cite{q11,q9}.
 Due to a $3m$ point-group symmetry for monolayer $\mathrm{SrAlGaSe_4}$,
 the  piezoelectric stress   and strain tensors by using  Voigt notation  can be reduced into :
 \begin{equation}\label{pe1-1-1}
 e=\left(
    \begin{array}{ccc}
      e_{11} & -e_{11} & 0 \\
     0 & 0 & -e_{11} \\
      e_{31} & e_{31} & 0 \\
    \end{array}
  \right)
    \end{equation}

  \begin{equation}\label{pe1-2-2}
  d= \left(
    \begin{array}{ccc}
      d_{11} & -d_{11} & 0 \\
      0 & 0 & -2d_{11} \\
      d_{31} & d_{31} &0 \\
    \end{array}
  \right)
\end{equation}
When   a uniaxial in-plane strain is applied,  monolayer $\mathrm{SrAlGaSe_4}$ has both in-plane and vertical piezoelectric polarization ($e_{11}$/$d_{11}$$\neq$0 and $e_{31}$/$d_{31}$$\neq$0).
However, when the  biaxial in-plane strain is applied,  the
in-plane piezoelectric response will be suppressed, while the
out-of-plane one still will remain ($e_{11}$/$d_{11}$=0 and $e_{31}$/$d_{31}$$\neq$0).
The $e_{ij}$ can be calculated by DFPT, and the
$d_{ij}$  can be  derived by  \autoref{pe1-4}, \autoref{pe0-1-1}, \autoref{pe1-1-1} and \autoref{pe1-2-2}:
\begin{equation}\label{pe2-2}
    d_{11}=\frac{e_{11}}{C_{11}-C_{12}}~~~and~~~d_{31}=\frac{e_{31}}{C_{11}+C_{12}}
\end{equation}

Here, we investigate the piezoelectric properties  of monolayer $\mathrm{SrAlGaSe_4}$ with 1.06 strain as a representative TI.
Firstly,  the $C_{11}$ and $C_{12}$ are calculated with GGA by SSR, and they are 54.14 $\mathrm{Nm^{-1}}$ and 23.30 $\mathrm{Nm^{-1}}$, which are smaller than ones of unstrained $\mathrm{SrAlGaSe_4}$ monolayer. Tensile strain reduced $C_{11}$ and $C_{12}$ have been found in many 2D materials\cite{gsd1,gsd3}.
And then, we use the   orthorhombic supercell as the computational  cell (in \autoref{t0})  to calculate  $e_{ij}$  with GGA+SOC by DFPT.
The $e_{ij}$ of monolayer $\mathrm{SrAlGaSe_4}$ are calculated, and
the calculated in-plane $e_{11}$ and out-of-plane  $e_{31}$ are -0.575$\times$$10^{-10}$ C/m and -0.053$\times$$10^{-10}$ C/m.
Based on \autoref{pe2-2}, the calculated $d_{11}$ and $d_{31}$ are -1.865 pm/V and -0.068 pm/V. So, the monolayer $\mathrm{SrAlGaSe_4}$  can become a potential PQSHI by strain.

\begin{figure}
   \includegraphics[width=8.0cm]{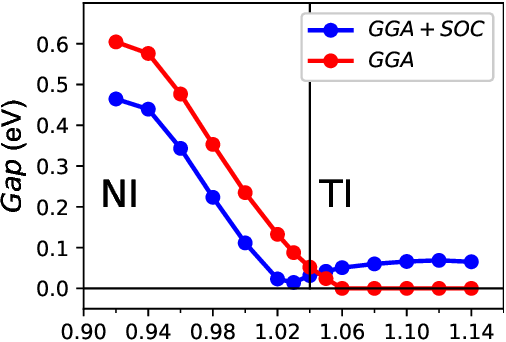}
  \caption{(Color online) The energy band gap of $\mathrm{CaSiGeN_4}$ monolayer as a function of  $a/a_0$ by using both GGA and GGA+SOC.}\label{t3-1-1}
\end{figure}

\section{Discussions and Conclusion}
 For $\mathrm{MoSi_2N_4}$ monolayer, the HSE06 (GGA) gives an indirect gap semiconductor with the gap of 2.297 (1.744) eV, and the experimental value is 1.94 eV\cite{msn}. The difference between HSE06 (GGA) and experimental value is 0.357 (-0.196) eV. So, it may be more suitable for  $\mathrm{MA_2Z4}$ family to use GGA to study their electronic properties.
Although the GGA  may  underestimate energy gap of  monolayer  $\mathrm{SrAlGaSe_4}$,  our predicted PQSHI should be  qualitatively correct, and only the critical point of NI to TI phase transition changes. The focus of our works is to provide a idea to achieve PQSHI , and many PQSHIs should be constructed in  $\mathrm{MA_2Z4}$ family. To confirm that, we also investigate the NI to TI phase transition of $\mathrm{CaAlGaSe_4}$ monolayer caused by strain. The optimized lattice constants is 4.02 $\mathrm{{\AA}}$, and the calculated $C_{11}$ and $C_{12}$ are  89.95 $\mathrm{Nm^{-1}}$ and  30.71 $\mathrm{Nm^{-1}}$, which satisfy the  Born  criteria of mechanical stability\cite{ela}.  The dynamical stability is also proved by phonon band dispersions  of  $\mathrm{CaAlGaSe_4}$ monolayer from Fig.1 of electronic supplementary information (ESI), and the thermal  stability is confirmed from Fig.2 of ESI.
 The energy band gaps of both GGA and GGA+SOC as $a$/$a_0$ function are shown in \autoref{t3-1-1}, and the energy bands at representative strain points are plotted in Fig.3 of ESI. It is found that the transition point of NI to TI is about 1.04, which is larger than one of $\mathrm{SrAlGaSe_4}$. The evolution of WCC and edge states of $\mathrm{CaAlGaSe_4}$ monolayer at  representative 1.06 strain are shown in Fig.4 and Fig.5 of ESI, which clearly show the nontrivial band topology.
At representative 1.06 strain, the calculated in-plane $e_{11}$ and out-of-plane  $e_{31}$ are 3.071$\times$$10^{-10}$ C/m and -0.066$\times$$10^{-10}$ C/m.
Based on \autoref{pe2-2}, the calculated $d_{11}$ and $d_{31}$ are 8.07 pm/V and -0.077pm/V with $C_{11}$ and $C_{12}$ being  61.59 $\mathrm{Nm^{-1}}$ and  23.52 $\mathrm{Nm^{-1}}$. These show that $\mathrm{CaAlGaSe_4}$ monolayer can become PQSHI by tensile strain.

In summary,  our DFT calculations demonstrate that it is
possible to realize  piezoelectricity and robust nontrivial band topology in a single material. The electronic structure of Janus monolayer  $\mathrm{SrAlGaSe_4}$ with piezoelectric properties at very small strain can act as a prototype for designing PQSHI.
Moreover, the PQSHI can coexist with Rashba spin splitting due to lacking vertical reflection symmetry.
Most importantly, many PQSHIs can be achieved in 2D  $\mathrm{MA_2Z_4}$ family by using the same design principle of $\mathrm{SrAlGaSe_4}$ monolayer, for example Janus $\mathrm{CaAlGaSe_4}$ monolayer.
 The  realization of PQSHI with  Rashba spin splitting can potentially lead to new device applications in electronics and spintronics, and can stimulate further studies for multifunctional 2D materials.

\begin{acknowledgments}
This work is supported by the Natural Science Foundation of Shaanxi Provincial Department of Education (19JK0809). We are grateful to the Advanced Analysis and Computation Center of China University of Mining and Technology (CUMT) for the award of CPU hours and WIEN2k/VASP software to accomplish this work.
\end{acknowledgments}

\end{document}